\documentclass[3p,twocolumn]{elsarticle}

\usepackage{lineno,hyperref,amsmath,amssymb}

\journal{Nuclear Instruments and Methods in Physics Research Section B}

\newcommand{\fwhmpct}{10\% }
\newcommand{\epeakpct}{1\% }

\begin{document}

\begin{frontmatter}

\title{Analytic saddlepoint approximation for ionization energy loss distributions}

\author{S.~K.~L Sjue\corref{corrauth}}
\cortext[corrauth]{Corresponding author}
\ead{sjue@lanl.gov}
\author{R.~N. George\corref{}}
\author{D.~G. Mathews\corref{}}
\address{Physics Division, Los Alamos National Laboratory, Los Alamos, NM 87545, USA}

\begin{abstract}
We present a saddlepoint approximation for ionization energy loss distributions, valid for arbitrary relativistic velocities of the incident particle $0 \leq v/c \leq 1$, provided that ionizing collisions are still the dominant energy loss mechanism.  We derive a closed form solution closely related to Moyal's distribution.  This distribution is intended for use in simulations with relatively low computational overhead.  The approximation generally reproduces the Vavilov most probable energy loss and full width at half maximum to better than \epeakpct and \fwhmpct, respectively, with significantly better agreement as Vavilov's $\kappa$ approaches 1.
\end{abstract}


\end{frontmatter}


\section{Introduction}

The motivation for this study is to provide a fast approximate method to generate energy loss distributions for heavy charged particles passing through thick, dense materials.  The primary goal is to improve the fidelity of our simulations for proton radiography \cite{sjueRSIpRad} by adding straggling distributions as a function of energy loss, without adding computational time penalty.  While there is a large body of work on this subject, much of which would be suitable for our application in one form or another, the main advantage of this study is a fast, compact solution which is suitable over a large range of energy losses for arbitrary values of $v/c$ for the incident particle.  

Our starting point is the approximation applied by Moyal \cite{moyal} to find a closed form approximation to the Landau distribution \cite{landau}.  It will be further developed to obtain a more general solution applicable to higher incident energies and larger energy losses like Vavilov's solutions \cite{vavilov}.  This further development of the saddlepoint approximation has been pursued elsewhere \cite{sigmund-winterbon}, but our generalization leads to a simpled closed form solution valid for arbitrary relativisitic velocities.  

\section{Energy loss formulae}

The single collision energy loss spectrum for large energy losses comes directly from the Rutherford \cite{rutherford} and Mott \cite{mott} differential cross sections in the nonrelativistic and relativistic cases, respectively.  Converting the Rutherford cross section to energy loss in cgs units (\textit{e.g.}, \cite{bichsel-fundamentals}), one finds
\begin{equation}
\left(\dfrac{d\sigma}{d\varepsilon}\right)_R = \dfrac{2\pi z^2 e^4}{m_e v^2 \varepsilon^2},
\end{equation}
where $z$ is the charge of the incident ionizing particle, $v$ is its velocity, $\varepsilon$ is the energy lost in the collision and $m_e$ is the mass of the electron.  Since we are primarily concerned with protons, the remainder of this article will assume $z=1$.  The same equation can be applied to energy loss in collisions with the nuclei in the stopping material with the substitution $m_e \rightarrow M(Z,A)$, showing that the energy loss in these interactions is orders of magnitude smaller.

The same conversion using the Mott cross section for relativistic incident particles gives
\begin{equation}
\left(\dfrac{d\sigma}{d\varepsilon}\right)_M = 
\left(\dfrac{d\sigma}{d\varepsilon}\right)_R 
[1-\beta^2 \varepsilon / T_m],
\end{equation}
where $\beta=v/c$ is the velocity of the incident particle in units of the speed of light and $T_m$ is the maximum kinematically allowed energy loss in a single collision.  Further corrections dependent on the spin of the incident particle are neglected since they are proportional to $\varepsilon/E$ or $m_e/M$, where $M$ is the incident particle's velocity.

The maximum kinematically allowed energy loss in a single collision, written in terms of the kinetic energy $T=E-M$, is
\begin{equation}
T_m=2m_e \dfrac{(T+M)^2-M^2}{M^2+2(T+M)m_e+m_e^2}.
\end{equation}
This reduces to $T_m\approx 2m_e \beta^2 \gamma^2$ for $(T+M)m_e/M^2 \ll 1$.  This value agrees with the particle data group (PDG) \cite{pdg}, but it uses kinetic energy instead of $\gamma$ and $\beta$ for reference.

The average energy loss is given by the Bethe-Bloch equation:
\begin{equation}
-\left\langle \dfrac{dE}{dx} \right\rangle = 
\dfrac{KZ\rho}{A\beta^2} 
\left[ \ln \dfrac{\sqrt{2 m_e \beta^2 \gamma^2 T_m}}{I} - \beta^2 \right].
\label{eq:bethe-bloch}
\end{equation}
The mean excitation energy $I$ accounts for the average effect of all the bound electron states.  The constant $K=4\pi N_A e^4/m_e=0.307075$ MeV cm$^2$/mol.  For compound materials, a suitable average can be made for $\langle Z/A \rangle$ and $\langle 1/I \rangle$, but we will assume an elemental material for simplicity.  We neglect the density correction $\delta$ \cite{sternheimer} in the following sections, but it can be included by replacing the second term in the square brackets with $\beta^2\rightarrow\beta^2+\delta/2$.

\section{Energy loss moments}

The moments of the total energy loss cross section determine the straggling distributions.  When $T_m\approx 2 m_e \beta^2 \gamma^2$ is a good approximation, the energy loss given by Equation \ref{eq:bethe-bloch} is equal to twice the energy loss given by $(d\sigma/d\varepsilon)_M$ with the minimum energy loss set to $I$.  We will calculate an adjusted value $I'$ to account for the difference between the assumed and true energy loss cross sections.  

Given the assumed form of the cross section $(d\sigma/d\varepsilon)_M$, the mean number of collisions $N$ after passage through a thickness $x$ is
\begin{align}
N= & x \dfrac{Z N_A \rho}{A} \int_{I'}^{T_m} \left(\dfrac{d\sigma}{d\varepsilon}\right)_M d\varepsilon \\
 = & \dfrac{K Z \rho x}{2A\beta^2} \left[\dfrac{1}{I'}-\dfrac{1}{T_m}\left(1+\beta^2 \ln \dfrac{T_m}{I'}\right)\right] \\
 = & \sigma_T \dfrac{Z N_A \rho x}{A}.
\end{align}
The last line defines the effective total cross section, $\sigma_T$; it is given by $K/2N_A\beta^2 I'$, if the term of order $I'/T_m$ is neglected.  Setting $N=1$ and solving for $x$ gives a value for the effective average distance between collisions,
\begin{equation}
x_0=\dfrac{1}{\sigma_T}\dfrac{A}{ZN_A \rho}.
\end{equation}

The mean energy loss will be taken as a given, determined by the Bethe-Bloch equation with any necessary corrections.  In terms of $I'$ and the assumed Mott form of the cross section it is
\begin{equation}
-\left\langle\dfrac{dE}{dx}\right\rangle = \dfrac{KZ\rho}{2A\beta^2} \left[ \ln \left( \dfrac{T_m}{I'} \right) - \beta^2 \right].
\end{equation}
This equation defines $I'$ for a given value of $\langle dE/dx \rangle$.  Then it is possible to get a precise value for the average energy loss while correctly capturing the high energy loss part of the cross section that is most relevant to straggling distributions with energy losses $\Delta > T_m$.  The solution for $I'$ in terms of Equation \ref{eq:bethe-bloch} is 
\begin{equation}
I'=Ie^{\beta^2}\dfrac{I}{2 m_e \beta^2 \gamma^2}.
\end{equation}
For high energy incident particles, the last factor on the right side will be very small and $I'$ will be smaller than any physical minimum energy loss.  The straggling distributions derived from these assumptions will be too narrow for small energy losses, when compared with realistic cross sections including minimum energy losses and resonant enhancement of the collision cross section with bound electrons.  Inclusion of the density correction $\delta$ changes the argument in the exponential to $\beta^2+\delta/2$.

The asymptotic width of the straggling distribution in a central limit theorem approximation is 
\begin{align}
\sigma^2(x)= & \dfrac{ZN_A\rho}{A}x\int_{I'}^{T_m} \varepsilon^2 \left(\dfrac{d\sigma}{d\varepsilon}\right)d\varepsilon \\
           = & \dfrac{KZ\rho x}{2A\beta^2} \int_{I'}^{T_m}\left(1-\beta^2 \dfrac{\varepsilon}{T_m}\right) d\varepsilon \\
           \approx & \dfrac{K Z \rho x}{2A\beta^2} T_m (1-\beta^2/2),
\end{align}
which agrees with Gaussian approximations presented elsewhere \cite{seltzerberger,leroyprinciples}.  However, the Gaussian approximation is generally not satisfactory for moderate energy losses satisfying the assumption that $\beta$ and $d\sigma/d\varepsilon$ do not change.

\section{Saddlepoint approximation}

The objective is to find energy loss distributions for massive, energetic particles passing through thick objects.  For the sake of simple, analytic forms the treatment will be approximate.  The initial state of interest is a monochromatic beam of particles incident on a finite thickness of material.  

We begin with a version of of the Landau-Bothe equation \cite{landau,bothe,sigbook} written in the same fashion as Moyal \cite{moyal},
\begin{equation}
F(\Delta,N)=\dfrac{1}{2\pi i} 
\int_{s_0-i\infty}^{s_0+i\infty}
ds \exp [s\Delta + N M(s)],
\label{eq:laplaceinversion}
\end{equation}
in which $\Delta$ is the energy loss, $N$ is the mean number of collisions, and $R(s)$ is a moment generating function given by
\begin{equation}
M(s)=\int_1^{\varepsilon_m} d\varepsilon
\dfrac{d\sigma}{d\varepsilon} e^{-s\varepsilon} -1.
\label{eq:mgfint}
\end{equation}
This cross section is normalized so that $F(\Delta,N)$ is written in terms of expected number of collisions instead of the equivalent thickness.  The integration limits are based on factorization of the minimum energy loss in a collision, $I'$, such that $\varepsilon_m=T_m/I'$.

Equation \ref{eq:laplaceinversion} is a standard Laplace inversion formula and the moment generating function in the exponential encapsulates the form of the cross section and fluctuations in the number of collisions.  For given energy loss $\Delta$ and expected number of collisions $N$, the integral for $F(N,\Delta)$ has a saddle point given by 
\begin{equation}
\Delta=-NM'(s).
\label{eq:spcondition}
\end{equation}
Taylor expansion of the argument to second order gives a Gaussian integral with amplitude
\begin{equation}
F(\Delta,N) \approx \dfrac{\exp\{N[M(s)-sM'(s)]\}}{\sqrt{2\pi N M''(s)}}.
\label{eq:spamp}
\end{equation}
The saddle point condition in Equation \ref{eq:spcondition} defines the energy loss as a function of $s$ and $F(\Delta,N)$ gives the amplitude.

The saddle point amplitude requires the moment generating function and its derivatives.  The integral in Equation \ref{eq:mgfint} can be written in terms of either the Rutherford or Mott forms of the cross section.  If $M(s)$ is used for the relativistic Mott form and $R(s)$ is the moment generating function for the Rutherford cross section, then we have
\begin{equation}
M(s)=R(s)+\dfrac{\beta^2}{\varepsilon_m}R'(s).
\end{equation}
Integration by parts gives
\begin{align}
R(s)=&e^{-s}-s\int_1^{\varepsilon_m}d\varepsilon \dfrac{e^{-s\varepsilon}}{s}-1 \\
    =&e^{-s}+sR'(s)-1.
\end{align}
The first derivative is related to the exponential integral,
\begin{equation}
R'(s)=-\int_1^{\varepsilon_m}d\varepsilon \dfrac{e^{-s\varepsilon}}{\varepsilon},
\end{equation}
while the second derivative takes a simple closed form:
\begin{equation}
R''(s)=\int_1^{\varepsilon_m} d\varepsilon e^{-s\varepsilon_m}
=\dfrac{e^{-s}-e^{-s\varepsilon_m}}{s}.
\end{equation}
The third derivative is
\begin{equation}
R^{(3)}(s)=-\dfrac{1}{s}\left[R''(s)+e^{-s}-\varepsilon_m e^{-s\varepsilon_m}\right],
\end{equation}
and for completeness the fourth derivative is
\begin{equation}
R^{(4)}(s)=-\dfrac{1}{s}\left[2R^{(3)}(s)-e^{-s}+\varepsilon_m^2e^{-s\varepsilon_m}\right].
\end{equation}
These derivatives are necessary to solve for the most probable energy loss using the relativistic form of the cross section.

\section{Moyal's solution}

Moyal gave a closed form solution based on the same limits originally studied by Landau.  The relevant limits are $s \ll 1 \ll s\varepsilon_m$, or equivalently, $N \gg 1$ and $\Delta \ll \varepsilon_m$.  In this limit, the leading order contribution to the second derivative is 
\begin{equation}
R''(s)\approx \dfrac{1}{s}-1+{\cal O}(s),
\end{equation}
which we integrate to find
\begin{equation}
R'(s) = \ln(s)+c.
\label{eq:rpmoyal}
\end{equation}
The precise integration constant is the Euler-Mascheroni constant $c=0.5772\ldots$  The energy loss in terms of $s$ is
\begin{equation}
\Delta=-N \ln(s)-Nc,
\end{equation}
 the value of $s$ in terms of the energy loss is
\begin{equation}
s = e^{-\Delta/N+c},
\end{equation}
and the leading dependence of $NR(s)-NsR'(s)$ is given by $-Ns$.  Putting it all together, the result is
\begin{align}
F(\Delta,N)\propto \sqrt{s} e^{-Ns}.
\end{align}
Setting the derivative equal to zero, the maximum value of $F(\Delta,N)$ is at $s=1/2N$ and the most probable energy loss is
\begin{equation}
\Delta_p=N\ln(2N)-Nc
\end{equation}
Using a reduced variable $\omega=(\Delta-\Delta_p)/N$ to write the solution in terms of the most probable energy loss, the result is Moyal's distribution,
\begin{equation}
\dfrac{dF}{d\omega}=\dfrac{1}{\sqrt{2\pi}} \exp\left\{-\dfrac{1}{2}[\omega+\exp(-\omega)]\right\},
\label{eq:moyaldist}
\end{equation}
where the normalization constant follows from a Gaussian integral upon the substitution $u=\exp(-\omega/2)$.

\begin{figure}
\includegraphics[width=0.5\textwidth]{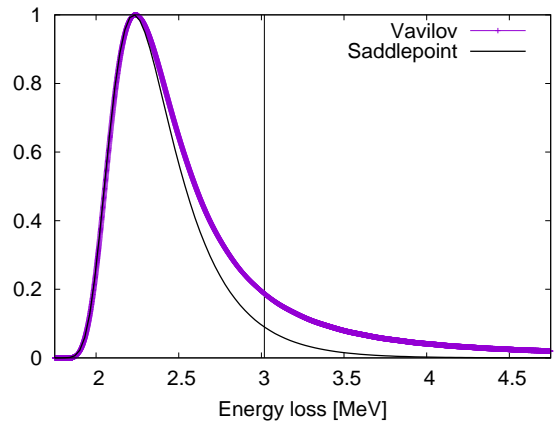}
\caption{Comparison between Vavilov and saddlepoint results for protons with 10 GeV kinetic energy passing through 2 mm of Pb, which corresponds to Vavilov's $\kappa=0.001$ or Landau's $\xi\approx 0.136$.  The average energy loss, marked by the vertical line, comes from the NIST PSTAR program \cite{pstar}.  The saddlepoint result has the parameter $t=0.501$, so this is essentially the Landau-Moyal limit.  This is the worst case for the saddlepoint approximation.  In this limit the discrepancy becomes severe at and above the mean energy loss.}
\label{fig:pbcomp}
\end{figure}

If we rewrite the variables, the result for the most probable energy loss is
\begin{equation}
\Delta_p=\xi \left( \ln \dfrac{\xi}{I'} + \ln 2 - c \right),
\end{equation}
where $\xi$ has the same definition originally given by Landau
\begin{equation}
\xi=x \dfrac{KZ\rho}{2A\beta^2}
=x\dfrac{2\pi e^4}{m_e\beta^2}\dfrac{\rho N_A Z}{A}.
\end{equation}
The peak energy loss grows like $x\ln x$ for small thicknesses; clearly, this behavior must change as $\Delta_p$ approaches the average energy loss.  The full width at half maximum (FWHM) of Moyal's distribution is $3.6\xi$, compared to a FWHM of $4\xi$ for Landau's distribution.  This reflects the limitation of the saddle point approximation and the first order treatment of $R'(s)$, but this is a worst case scenario in comparison to the Vavilov distribution, as we will show later.  Figure \ref{fig:pbcomp} is a good illustration of the difference between these two distributions.  Although the saddlepoint distribution has been calculated using the generalized form of the next section, it is effectively identical to the Moyal result in this case.

\section{Generalized form}

A more general form is necessary for higher energy ionizing particles and larger energy losses.  We seek a compact closed form which is applicable in such a scenario.  Our starting point is an assumption analogous to Equation \ref{eq:rpmoyal}, which satisfies three derivatives with respect to $s$, about the most probable energy loss, or equivalently the most probable value of $s$.

The most probable value of $s$ is found by setting the derivative of Equation \ref{eq:spamp} to zero.  The result is
\begin{equation}
s_p=-\dfrac{M^{(3)}(s)}{2NM''(s)}
\end{equation}
In the non-relativistic limit $\beta^2=0$, $s_p$ varies from $1/2N$ in the Landau-Moyal limit to $1/4N$ in the large energy loss limit.  The general result for the minimum value of $s_p$ is 
\begin{equation}
\textrm{min}(s_p)=\dfrac{1}{2N}\dfrac{1/2-\beta^2/3}{(1-\beta^2/2)^2},
\label{eq:spmin}
\end{equation}
with a limiting minimum value of $s_p=1/3N$ as $\beta^2\rightarrow 1$.

Our generalization of Equation \ref{eq:rpmoyal} is
\begin{align}
M'(s) \approx  & m_1 + r m_2 \ln \left(1+ \dfrac{s-s_p}{r}\right),
\label{eq:m1gen}
\end{align}
in which $s_p$ is the most probable value of $s$, $m_1=M'(s_p)$, $m_2=M''(s_p)$, $r=m_2/m_3$ and $m_3=|M^{(3)}(s_p)|$.  This form ensures the correct value of $s_p$ and the first three derivatives at $s_p$.  It also reduces to Moyal's appproximation for $R'(s)$ in the Landau-Moyal limit.  When we invert the approximate form for $M'(s)$, the first two derivatives in terms of the energy loss are
\begin{align}
M'(s)     =       & -\Delta/N \textrm{ and}\\
M''(s)    \approx & m_2 e^{\omega}, \textrm{ where} \\
\omega    =       & \dfrac{\Delta + Nm_1}{N r m_2 } = \dfrac{\Delta-\Delta_p}{Nrm_2}.
\end{align}
$\Delta_p$ is the most probable energy loss and this value of $\omega$ also reproduces Moyal's reduced energy loss variable in the small energy loss limit.

Now we wish to find a distribution analogous to Equation \ref{eq:moyaldist} with more general applicability.  The more complicated cross section does not allow the same simple reduction of the exponent in the saddlepoint amplitude.  Instead we integrate the approximate form for $M'(s)$ to obtain $M(s)$,
\begin{align}
M(s)  =       & \int ds M'(s) \\
      \approx & m_1 s + r^2 m_2 \int du \ln u \\
      =       & m_1 s + r^2 m_2 (u \ln u - u ),
\end{align}
which has used the substitution $u=1+(s-s_p)/r= e^{-\omega}$.  It is fine to neglect a potential integration constant since we will renormalize the distribution obtained from the saddlepoint approximation in any case.  A simplified result for the exponent is
\begin{equation}
N(M-sM')=-r^2m_2N\left[\omega + e^{-\omega} \right]+\omega/2.
\end{equation}
The factor of $1/\sqrt{M''}$ cancels the last term since it takes the form $e^{-\omega/2}$.  The final renormalized distribution is
\begin{equation}
\dfrac{dF}{d\omega}=\dfrac{t^t}{\Gamma(t)}\exp\left\{-t[\omega+\exp(-\omega)]\right\},
\label{eq:RESULT}
\end{equation}
where we have defined
\begin{equation}
t=\dfrac{N m_2^3}{m_3^2}.
\end{equation}
If we integrate the distribution in terms of $te^{-\omega}$, it takes the integral form of the gamma function.  This gives the normalization constant.  The value of $t$ is 1/2 in the small energy loss limit, leading to Moyal's result.  Larger energy losses lead to larger values of $t$.  For values of $t\gg 1$, the Gaussian approximation is recovered if one expands the argument to second order: $\omega+e^{-\omega}\approx 1+\omega^2/2$.

\section{Efficient computations}

The primary difficulty with the outlined approach is to determine $s_p$ and $R'(s_p)$.  To this end we give simple methods to caclulate these values.  The condition for the most probable value of $s$ is 
\begin{equation}
y_p(s_p)=s_p-\dfrac{m_3}{2Nm_2^2}=0.
\end{equation}
In this equation we substitute $s_1=\textrm{min}(s_p)$ from Equation \ref{eq:spmin} and $s_2=\textrm{max}(s_p)=1/2N$ to find $y_1(s_1)=\delta s_1$ and $y_2(s_2)=\delta s_2$.  This is a monotonic function and it is linear in $s$ to an excellent approximation over the range in question.  Whichever value of $y$ is closer to zero, we iterate to the solution by setting $s(2)=s(1)-\delta s$.  This procedure rapidly converges to the precise value of $s_p$ and $\Delta_p$.

The higher derivatives have simple closed forms.  The precise value of the first derivative of the moment generating function is given by
\begin{equation}
R'(s)=E_1(s\varepsilon_m)-E_1(s),
\end{equation}
where $E_1$ is the exponential integral (see, for example, Abramowitz and Stegun \cite{abramowitz1966}).  In the Landau-Moyal limit only the second term is necessary.  It is useful to have quick solutions for this integral without special math libraries.  For values such that $s\varepsilon_m > 1$, a convenient solution is
\begin{equation}
R'(s)=\ln(s)+c-s+\dfrac{\exp(-s\varepsilon_m)}{s\varepsilon_m},
\end{equation}
which has truncated terms beginning with $s^2/4$ and $-\exp(-s\varepsilon_m)/(s\varepsilon_m)^2$.  A convenient formula for small values of $s\varepsilon_m$ is
\begin{equation}
R'(s)=\ln(\varepsilon_m)-\sum_{n=1}^{\infty} \dfrac{(-1)^n[(s\varepsilon_m)^n-s^n]}{nn!}.
\end{equation}
In this limit it will usually be acceptable to neglect $s^n$.  

\section{Comparisons with Vavilov distribution}

Now we demonstrate the performance of our approximate distribution relative to the well-known Vavilov distribution.  The results for the Vavilov distribution were generated using the ROOT package from CERN \cite{cernroot}; in particular, we have used the VavilovAccurate implementation of the programs by Schorr \cite{schorr1973}. 

The saddlepoint approximation reproduces the peak energy loss and distribution in the vicinity of the peak energy loss.  The approximation is worse for smaller $N$ and farther from the peak of the distribution.  To characterize the differences, we have compared the saddlepoint distributions to the Vavilov distribution as a function of Vavilov's $\kappa$, which is related to Landau's $\xi$ by
\begin{equation}
\kappa=\xi/T_m.
\end{equation}
The comparisons were made over the range $\kappa\in[0.001,10]$ for the values $\beta^2=\left\{0.1,0.5,0.9926\right\}$, corresponding to proton \textit{kinetic} energies of 50.8 MeV, 389 MeV and 10 GeV.  Over this range, the most probable energy loss is reproduced with discrepancies  less than 1\% in all cases.  The overall worst performance of the saddlepoint approximation is at small values of the energy loss or $\kappa$.  Figure \ref{fig:pbcomp} shows the energy loss of 10 GeV protons passing through about 2 mm of Pb; in this case the difference between the most probable energy losses is 0.15\%.  

\begin{figure}
\includegraphics[width=0.5\textwidth]{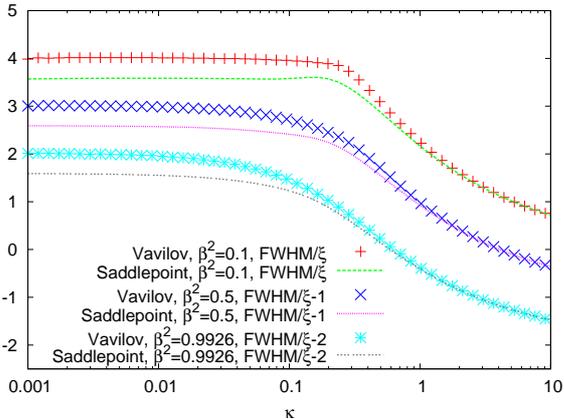}
\caption{Comparison between the Vavilov FWHM and the saddlepoint FWHM for $\beta^2=0.1$, 0.5, and 0.9926 as a function of $\kappa$.  The results have been offset on the vertical axis for visibility.
}
\label{fig:fwhmcomp}
\end{figure}

The FWHM illustrates the limitations of the saddlepoint method.  In the Moyal-Landau limit, the Landau distribution has FWHM$=4.02\xi$ and the Moyal distribution has FWHM=$3.59\xi$, so the Moyal distribution is 10.7\% too narrow by this metric.  It is worth noting that the Landau distribution is generally deficient in this limit itself: straggling distributions in this limit are wider due to enhanced cross sections at energy losses comparable with electron binding energies.  In all cases, the comparison between the Vavilov and saddlepoint FWHM improves with increasing energy loss.  This is shown for three values of $\beta^2$ in Figure \ref{fig:fwhmcomp}.  The difference in the FWHM at small energy losses is characteristic of the Landau and Moyal distributions.  The disagreement decreases with further energy loss.  At $\kappa=1$ the discrepancy has decreased to $<3\%$ for all three cases; the largest difference is 2.4\% for $\beta^2=0.1$ at $\kappa=1$.

\begin{figure}
\includegraphics[width=0.5\textwidth]{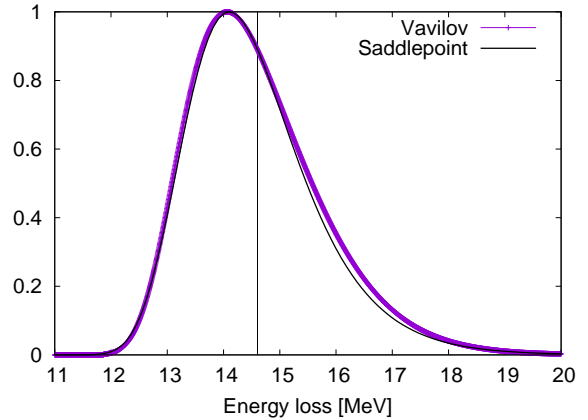}
\caption{Comparison between Vavilov and saddlepoint results for protons with 800 MeV kinetic energy passing through 1 cm of Cu, which corresponds to Vavilov's $\kappa\approx0.36$ or Landau's $\xi\approx 0.89$.  The average energy loss is marked by the vertical line.  The saddlepoint result has the parameter $t\approx1.4$.}
\label{fig:cucomp}
\end{figure}

For completeness, we include two more comparisons.  The mean energy losses in these comparisons were obtained using the PDG values for the average excitation energy and no other corrections were included.  Since we are interested in proton radiography with kinetic energy 800 MeV at the Los Alamos Neutron Science Center, we include protons at this energy ($\beta^2=0.7086$) passing through 1 cm of Cu in Figure \ref{fig:cucomp}.  The saddlepoint approximation with $t=1.4$ performs well at and above the mean energy loss in this case, which is a result of using Equation \ref{eq:m1gen} to get the moments correct at the peak energy loss.  The difference between the peak energy losses is about 40 keV with a peak energy loss of 14 MeV, or less than 0.3\%.  The FWHM is about 5\% smaller for the saddlepoint distribution.

\begin{figure}
\includegraphics[width=0.5\textwidth]{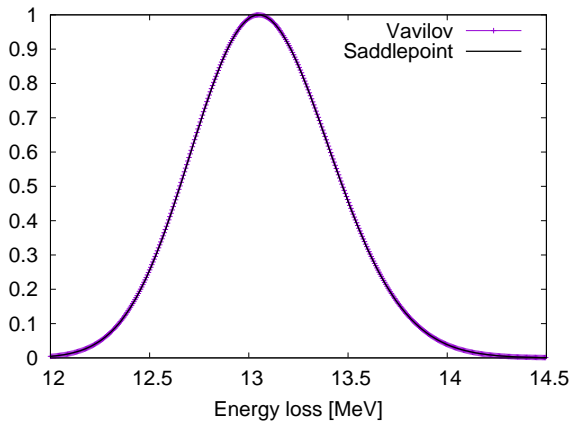}
\caption{Comparison between Vavilov and saddlepoint results for protons with $\beta^2=0.1$ and $\kappa=10$.  These energy losses correspond to thickness $\approx 1.8$ mm of Cu.}
\label{fig:kappa10}
\end{figure}

Finally, to show a case with a relatively large energy loss where the approximation performs very well, Figure \ref{fig:kappa10} shows distributions for protons with $\beta^2=0.1$ and Vavilov $\kappa=10$.  The distributions are indistinguishable on a linear scale.  Even for $\kappa=10$, the distribution is not symmetric.  The saddlepoint approximation is much better than a symmetric Gaussian.  An additional advantage is that saddlepoint approximation has a simpler functional form than a skewed Gaussian.

\section{Summary}

We have given a closed form saddlepoint approximation to Vavilov's distributions for the ionization energy loss of energetic charged particles.  This generally applicable distribution in Equation \ref{eq:RESULT} take the form of a simple extension to Moyal's closed form approximation for the Landau distribution.  The limit where the distribution takes the Moyal form is where it performs worst in comparison with the Landau-Vavilov forms.  Our approximate distribution correctly reproduces the most probable energy loss to better than 1\% over orders of magnitude in energy loss.  The width of the approximate distribution is about 10\% narrower than the Landau distribution in the thin limit and ultimately obtains the same form as the Vavilov distribution.  In practice, the approximate distribution is nearly identical to the Vavilov result for $\kappa \geq 1$.

We have described how to compute our distribution.  No special functions are required, only exponentials and logarithms.  The simple closed form of distribution has advantages.  The normalization constant only requires the calculation of $t^t/\Gamma(t)$.  The cumulative distribution function resulting from Equation \ref{eq:RESULT} can be expressed in terms of common special functions such as the incomplete gamma function and the exponential integral.  We find the performance of this approximate method more than adequate for our application to proton radiography.  We hope it might find broader application.

\section*{Acknowledgement}

This work was supported by the Advanced Radiography Science Campaign (C3) at Los Alamos National Laboratory.


\end{document}